\documentstyle[12pt,epsfig]{article}

\advance\textheight by \topskip
\begin{document}
\tolerance=100000
\thispagestyle{empty}
\setcounter{page}{0}

\newcommand{\be}{\begin{equation}}
\newcommand{\ee}{\end{equation}}
\newcommand{\br}{\begin{eqnarray}}
\newcommand{\er}{\end{eqnarray}}
\newcommand{\ba}{\begin{array}}
\newcommand{\ea}{\end{array}}
\newcommand{\bi}{\begin{itemize}}
\newcommand{\ei}{\end{itemize}}
\newcommand{\bn}{\begin{enumerate}}
\newcommand{\en}{\end{enumerate}}
\newcommand{\bc}{\begin{center}}
\newcommand{\ec}{\end{center}}
\newcommand{\ul}{\underline}
\newcommand{\ol}{\overline}
\def\epem{\ifmmode{e^+ e^-} \else{$e^+ e^-$} \fi}
\newcommand{\eeww}{$e^+e^-\rightarrow W^+ W^-$}
\newcommand{\qqQQ}{$q_1\bar q_2 Q_3\bar Q_4$}
\newcommand{\eeqqQQ}{$e^+e^-\rightarrow q_1\bar q_2 Q_3\bar Q_4$}
\newcommand{\eewwqqqq}{$e^+e^-\rightarrow W^+ W^-\ar q\bar q Q\bar Q$}
\newcommand{\eeqqgg}{$e^+e^-\rightarrow q\bar q gg$}
\newcommand{\eeqloop}{$e^+e^-\rightarrow q\bar q gg$ via quark loops}
\newcommand{\eeqqqq}{$e^+e^-\rightarrow q\bar q Q\bar Q$}
\newcommand{\eewwjjjj}{$e^+e^-\rightarrow W^+ W^-\rightarrow 4~{\rm{jet}}$}
\newcommand{\eeqqggjjjj}{$e^+e^-\rightarrow q\bar 
q gg\rightarrow 4~{\rm{jet}}$}
\newcommand{\eeqloopjjjj}{$e^+e^-\rightarrow q\bar 
q gg\rightarrow 4~{\rm{jet}}$ via quark loops}
\newcommand{\eeqqqqjjjj}{$e^+e^-\rightarrow q\bar q Q\bar Q\rightarrow
4~{\rm{jet}}$}
\newcommand{\eejjjj}{$e^+e^-\rightarrow 4~{\rm{jet}}$}
\newcommand{\jjjj}{$4~{\rm{jet}}$}
\newcommand{\qqbar}{$q\bar q$}
\newcommand{\ww}{$W^+W^-$}
\newcommand{\ar}{\rightarrow}
\newcommand{\sm}{${\cal {SM}}$}
\newcommand{\Dir}{\kern -6.4pt\Big{/}}
\newcommand{\Dirin}{\kern -10.4pt\Big{/}\kern 4.4pt}
\newcommand{\DDir}{\kern -7.6pt\Big{/}}
\newcommand{\DGir}{\kern -6.0pt\Big{/}}
\newcommand{\wwqqqq}{$W^+ W^-\ar q\bar q Q\bar Q$}
\newcommand{\qqgg}{$q\bar q gg$}
\newcommand{\qloop}{$q\bar q gg$ via quark loops}
\newcommand{\qqqq}{$q\bar q Q\bar Q$}
\newcommand{\ord}{{\cal O}}
\newcommand{\Ecm}{E_{\mathrm{cm}}}

\def\l{\left\langle}
\def\r{\right\rangle}
\def\aem{\alpha_{\rm em}}
\def\as{\alpha_{\rm s}}
\def\MW{M_{W^\pm}}
\def\MZ{M_{Z}}
\def\ycut{y_{\rm cut}}
\def\Ord{\lower .7ex\hbox{$\;\stackrel{\textstyle <}{\sim}\;$}}
\def\OOrd{\lower .7ex\hbox{$\;\stackrel{\textstyle >}{\sim}\;$}}
\def\pl #1 #2 #3 {{\it Phys.~Lett.} {\bf#1} (#2) #3}
\def\np #1 #2 #3 {{\it Nucl.~Phys.} {\bf#1} (#2) #3}
\def\jp #1 #2 #3 {{\it J.~Phys.} {\bf#1} (#2) #3}
\def\zp #1 #2 #3 {{\it Z.~Phys.} {\bf#1} (#2) #3}
\def\pr #1 #2 #3 {{\it Phys.~Rev.} {\bf#1} (#2) #3}
\def\prep #1 #2 #3 {{\it Phys.~Rep.} {\bf#1} (#2) #3}
\def\prl #1 #2 #3 {{\it Phys.~Rev.~Lett.} {\bf#1} (#2) #3}
\def\mpl #1 #2 #3 {{\it Mod.~Phys.~Lett.} {\bf#1} (#2) #3}
\def\rmp #1 #2 #3 {{\it Rev. Mod. Phys.} {\bf#1} (#2) #3}
\def\sjnp #1 #2 #3 {{\it Sov. J. Nucl. Phys.} {\bf#1} (#2) #3}
\def\cpc #1 #2 #3 {{\it Comp. Phys. Commun.} {\bf#1} (#2) #3}
\def\xx #1 #2 #3 {{\bf#1}, (#2) #3}
\def\preprint{{\it preprint}}

\begin{flushright}
{\large RAL-TR-99-032}\\
{\rm May 1999\hspace*{.5 truecm}}\\
\end{flushright}

\vspace*{\fill}

\begin{center}
{\Large \bf Can the trilinear Higgs self-coupling be measured\\[0.5cm]
at future linear colliders?}\\[1.cm]
{\large 
D.J. Miller\footnote{Email: D.J.Miller@rl.ac.uk}
and 
S.~Moretti\footnote{Email: Moretti@v2.rl.ac.uk}}\\[0.4 cm]
{\it Rutherford Appleton Laboratory,}\\
{\it Chilton, Didcot, Oxon OX11 0QX, UK.}\\
\end{center}

\vspace*{\fill}

\begin{abstract}
{\small
\noindent
We assess the possibility of measuring the trilinear self-coupling
of the Standard Model Higgs boson 
at future electron-positron linear colliders with centre-of-mass
energies between 500 and 1500 GeV. We consider rather light Higgs
scalars, with mass below the $W^+W^-$ threshold, produced in the double
Higgs-strahlung channel $e^+e^-\to HHZ$ and decaying via the dominant
mode $H\ar b\bar b$. Assuming ideal vertex tagging of the heavy-quark
jets and mass reconstruction of the $Z$ boson produced in association with
the Higgs pair, we compare the yield of the above process 
to those of all irreducible electroweak and strong backgrounds 
proceeding through a $b\bar bb\bar b Z$ intermediate state. 
Total cross sections and differential spectra of phenomenological
relevance to the selection and analysis 
of the signal are given and discussed.}
\end{abstract}

\vspace*{\fill}
\newpage

\section{Introduction}
\label{sec_intro}

In all probability a Higgs boson will be discovered at either LEP {\rm II}, 
Run {\rm II} of the Tevatron, or the Large Hadron Collider (LHC). It is then 
inevitable that the emphasis of Higgs physics will be turned away from
discovery and instead will focus on the investigation of Higgs boson
properties, such as its mass, width and branching ratios. Although much 
interesting Higgs phenomenology can be done at the LHC, many analyses
are made infeasible by the rather messy nature of hadron colliders. Instead
one must resort 
to the much cleaner environment of $e^+e^-$ annihilations, for
example, at the Linear Collider (LC)
\cite{ee500}, where precision measurements at the TeV scale can be made.

One particularly interesting task to be carried out at future colliders
is the reconstruction of the Higgs potential
itself, possibly confirming, or denying, the mechanism
of spontaneous electroweak symmetry breaking. This can be achieved by
measuring the trilinear $\lambda_{HHH}$ 
and quadrilinear $\lambda_{HHHH}$ Higgs self-couplings, which
can then be compared with the predictions of the Standard Model
(SM), or indeed the Minimal
Supersymmetric Standard Model (MSSM)\footnote{In principle, the former 
coupling is amenable to investigation also at hadron and high energy photon
colliders too, via double Higgs-strahlung off $W^\pm$ or $Z$ 
bosons \cite{revZHH,IPKSK},
$W^+W^-$ or $ZZ$ fusion \cite{IPKSK,BC,revWW},
gluon-gluon fusion \cite{revgg} or $\gamma \gamma$ fusion
\cite{IPKSK,BC,jikia}.}.

A measurement of  the trilinear term, $\lambda_{HHH}$, is the first step in 
reconstructing the Higgs potential. At a future $e^+e^-$ collider,  the 
$\lambda_{HHH}$ coupling of the SM is accessible through double 
Higgs-strahlung off $Z$ bosons, in the process $e^+e^-\to HHZ$. This is the 
mechanism with which we will be concerned in this paper (for the MSSM see 
Ref.~\cite{revMSSM,PMZ}). The SM signal of the trilinear Higgs 
self-coupling has been thoroughly investigated in Ref.~\cite{PMZ} (with its 
MSSM counterparts), and was found to be small but measurable for an 
intermediate mass Higgs boson, given a high integrated luminosity. 
In contrast, the quadrilinear vertex, $\lambda_{HHHH}$, is unmeasurable at 
the energy scale of the proposed LCs due to suppression by an additional 
power of the electromagnetic coupling constant. 

However, in measuring $\lambda_{HHH}$, one must be sure that the already 
small signal can be
distinguished from its backgrounds without being appreciably reduced. Here we
will examine the $H \to b\bar b$ decay channel over the Higgs mass range
$M_H\Ord140$ GeV and present kinematic cuts to aid its selection. The case of
off- and on-shell 
$H \to W^{\pm (*)}W^{\mp}$ decays for $M_H\OOrd140$ GeV is under
examination elsewhere \cite{jan}.

\begin{figure}[p]
~\epsfig{file=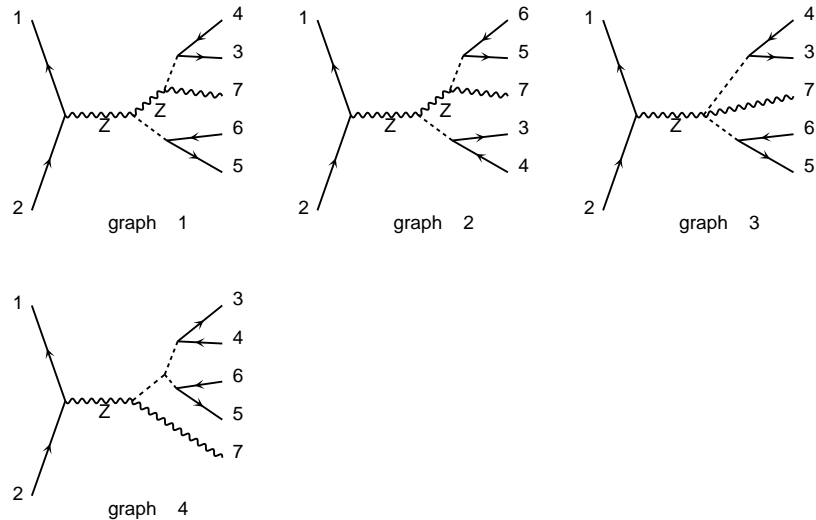,height=20cm}
\vskip-12.5cm
\caption{Diagrams contributing at lowest-order to
$e^+_1e^-_2 \to b_3\bar b_4 b_5\bar b_6 Z_7$ 
via purely EW interactions
containing two Higgs bosons in intermediate states. An internal wavy line
corresponds to a $Z$ boson (labeled as {\tt Z}).
The total number of actual diagrams is 4. Finally,
diagrams which differ from 
those above only in the exchange $3\leftrightarrow5$ 
(or, equivalently, $4\leftrightarrow6$) must also be considered,
preceded by a minus sign.}
\label{fig_HH}
\end{figure}

If one assumes very efficient tagging and high-purity sampling of $b$
quarks, the backgrounds to a $\lambda_{HHH}$ measurement from double
Higgs events in the $4b$ decay channel are primarily the `irreducible' 
ones via $b\bar b b\bar b Z$ intermediate states \cite{Lutz}, which can be 
separated into EW and QCD backgrounds. Furthermore, the double Higgs-strahlung 
process (see Fig.~\ref{fig_HH}):
\begin{equation}\label{eeHHZ}
e^+e^- \to HHZ\ar b\bar b b\bar b Z
\end{equation}
contains diagrams proceeding via an $HHZ$ intermediate state but not dependent 
on $\lambda_{HHH}$ (graphs 1--3 in Fig.~\ref{fig_HH}), as well as that 
sensitive to the trilinear Higgs self-coupling (graph 4 in Fig.~\ref{fig_HH}).
In addition, we also include four extra diagrams, which differ
only in the exchange of the four-momenta and helicities of two
identical $b$ quarks (or, equivalently, antiquarks) and a minus sign
(due to Fermi-Dirac statistics pertinent to identical fermionic particles).
However, the narrow width of the Higgs resonance ensures that the interference
will be negligible and these extra diagrams could be included by symmetry.

The other two backgrounds proceed via purely EW interactions 
(see Figs.~\ref{fig_H}--\ref{fig_N}),
\begin{equation}\label{eeEW}
e^+e^- \to~{\mathrm{EW~graphs}}~\ar b\bar b b\bar b Z,
\end{equation}
and via QCD couplings as well (see Fig.~\ref{fig_Q}),
\begin{equation}\label{eeQCD}
e^+e^- \to~{\mathrm{QCD~graphs}}~\ar b\bar b b\bar b Z,
\end{equation}
and both contain no more than one intermediate Higgs boson.
The EW background, process~(\ref{eeEW}), is of $\ord(\alpha_{em}^5)$
away from resonances, but can, in principle, be problematic due to the
presence of both $Z$ vectors and Higgs scalars yielding $b \bar b$ pairs.
Finally, the QCD background, process~(\ref{eeQCD}), is of $\ord(\alpha_{em}^3
\alpha_s^2)$ away from resonances. Here, although there are no
heavy objects decaying to
$b \bar b$ pairs, the production rate itself could give
difficulties due to the presence of the strong coupling. As with 
process~(\ref{eeHHZ}),
one must include diagrams with the interchange of the two identical $b$
(anti)quarks also in the EW and QCD background processes.
In contrast to the signal, here interference effects are sizable.

The plan of the paper is as follows. The next Section  details
the procedure adopted in computing the relevant scattering amplitudes.
Sect.~\ref{sec_results} displays our numerical results and contains our
discussion. Finally, in the last Section, we summarize and conclude.

\section{The matrix elements (MEs)}
\label{sec_calculation}

The double Higgs-strahlung process (\ref{eeHHZ}) proceeds at 
lowest-order through
the diagrams of Fig.~\ref{fig_HH},  as
explained in the Introduction.
They are rather straightforward 
to calculate in the case of on-shell $HHZ$ production
(see, e.g., Ref.~\cite{PMZ} for an analytic expression of the ME).

The EW background (\ref{eeEW}) derives from many
graphs: 550 in total (again, considering the $b$ (anti)quark statistics).
However, they can conveniently be grouped into different `topologies':
that is, collections of diagrams with identical (non-)resonant structure.
We have isolated 23 of these, and displayed them in Figs.~\ref{fig_H}
and \ref{fig_N}, depending on whether one or zero Higgs intermediate
states are involved, respectively. There are 214 graphs of the first kind
and 336 of the second.
This approach, of splitting the ME in (non-)resonant subprocesses,
 facilitates the integration over the phase space and 
further provides an insight into the 
fundamental dynamics. On the one hand, one can compute each of the 
topologies separately, with the appropriate mapping of variables,
thus optimizing the accuracy of the numerical integration.
On the other hand, one is able to assess the relative weight of the 
various subprocesses into the full scattering amplitude, 
by comparing the various integrals with each other. However, one should recall
that the amplitudes squared associated to each of these topologies
are in general non-gauge invariant. In fact, the latter is recovered only when 
the various (non)-resonant terms are summed up. For reasons of space, 
we will not dwell in technicalities any further here, as a good guide
to this technique can be found in Ref.~\cite{paps}. 
(The resonant structure of the various subchannels ought to be self-evident
in Figs.~\ref{fig_H}--\ref{fig_N}.)

The QCD diagrams associated to process (\ref{eeQCD}) can be found
in Fig.~\ref{fig_Q}. In total, one has 120 of these, with only five
different (non-)resonant topologies. 
The integration in this case is much simpler
than in the EW case and can in fact be carried out with percent accuracy
directly over the full ME using standard multichannel Monte Carlo methods.

Non-zero interference effects exist between processes (\ref{eeHHZ}), 
(\ref{eeEW}) and (\ref{eeQCD}). However, given the very
narrow width of the Higgs boson (always below 20 MeV over the mass range 
considered here), any interference with the signal can be safely neglected.
Furthermore, we will see that the dominant subprocesses of the two backgrounds have very different topologies,
so one also expects their interference to be negligible. Therefore,
given that their calculation would be rather cumbersome,
we do not consider them in our analysis. 

The large number of amplitudes can easily and efficiently be dealt with
in the numerical evaluation if one resorts to helicity amplitudes.
In doing so, we have adopted the {\tt HELAS} subroutines \cite{HELAS}.
The algorithm used to perform the multi-dimensional integrations
was {\tt VEGAS} \cite{VEGAS}.
 
Numerical inputs were as follows. 
The strong coupling constant $\alpha_{s}$ entering the QCD process 
(\ref{eeQCD}) has been evaluated at two loops, with $N_f=5$ and 
$\Lambda_{\overline{\mathrm {MS}}}=160$ MeV,
at a scale equal to the collider CM energy, $\sqrt s\equiv E_{\mathrm{cm}}$.
The EM coupling constant was  $\alpha_{{em}}=1/128$.
The sine squared of the Weinberg angle was $\sin^2\theta_W=0.232$.
The fermionic (pole) masses were $m_e=0$ and $m_b=4.25~{\mathrm{GeV}}$. 
As for the gauge boson masses (and their widths), we have used:
$M_Z=91.19~{\mathrm {GeV}}$, $\Gamma_Z=2.50~{\mathrm {GeV}}$,
$M_W\equiv M_Z\cos\theta_W \approx80~{\mathrm {GeV}}$ and 
$\Gamma_W=2.08~{\mathrm {GeV}}$.

Concerning the Higgs boson, we have spanned its mass $M_H$ over the
range 100 to 150 GeV and
we have computed its width, $\Gamma_H$, by means of the program
described in Ref.~\cite{WJSZK}, which  uses 
a running $b$ mass in evaluating the $H\ar b\bar b$ decay fraction. Thus,
for consistency, we have evolved  the value of $m_b$ entering the
$Hbb$ Yukawa coupling of the $H\ar b\bar b$ decay currents  
in the same way. 

We have adopted as CM energies typical for the LC the values
$E_{\mathrm{cm}}=500, 1000$ and 1500 GeV.

Notice that, in the remainder of this paper, 
total and differential rates are those 
at the partonic level, as we identify
jets with the partons from which they originate. 
In order to resolve the latter as four separate systems, we impose
the following acceptance cuts: $E(b)>10$ GeV on the energy of
each $b$ (anti)quark and $\cos(b,b)<0.95$ on the relative separation
of all possible $2b$ combinations. We further
assume that $b$ jets are distinguishable
from light-quark and gluon jets (e.g., by using $\mu$-vertex tagging 
techniques). However,
no efficiency  to tag the four $b$ quarks is included in our results.
Also, the $Z$ boson is treated as on-shell and no branching ratio
(BR) is applied to 
quantify its possible decays. In practise, 
in order to simplify the 
treatment of the final state,
one may assume that the $Z$ boson decays leptonically 
(i.e., $Z\to \ell^+\ell^-$, with $\ell=e,\mu,\tau$)
or hadronically into
light quark jets (i.e., $Z\to q\bar q$, with $q\ne b$).

Finally, we have not included Initial State Radiation
(ISR) \cite{ISR} in our calculations. 
In fact, we would expect it to affect rather similarly the
various processes (\ref{eeHHZ})--(\ref{eeQCD}). As we are basically interested
in relative rates among the latter, we are confident that the salient
features of our results are indifferent to the presence or not
of photons radiated by the incoming electron-positron beams\footnote{We
also neglect beamsstrahlung and Linac energy
spread, by assuming a narrow beam design \cite{ISR}.}. 
 
\section{Results}
\label{sec_results}

The total cross sections for process (\ref{eeHHZ}), at the three
CM energies considered here, can be found in the top-left
frame of Fig.~\ref{fig_cross}, as a function of $M_H$. 
The decrease of the total rates 
with increasing Higgs mass is mainly the effect of the BR
of the decay channel $H\to b\bar b$, see, e.g., Fig.~1 of Ref.~\cite{WJSZK}.
This mode is dominant and very close to 1 up to the opening
of the off-shell $H\ar W^{\pm*}W^{\mp}$ decay channel, which occurs
at $M_H\approx140$ GeV. 
In contrast, the production cross section for $e^+e^-\to HHZ$
is much less sensitive to $M_H$ \cite{PMZ}.
In addition, because reaction (\ref{eeHHZ}) is an annihilation process
proportional to $1/s$, a larger CM energy tends to deplete the production
rates, as long as $E_{\mathrm{cm}}\gg 2M_H+M_Z$. When this is no longer
true, e.g., at 500 GeV and $M_H\OOrd140$ GeV, phase space
suppression can overturn the $1/s$ propagator effects. This is evidenced by
the crossing of the curves for 500 and 1000 GeV.

In practice, the maximum cross section for double Higgs-strahlung
(\ref{eeHHZ}) is reached at energies $E_{\mathrm{cm}}\approx2M_H+M_Z+
200$ GeV \cite{PMZ}. For Higgs masses in the lower part of the 
$M_H$ range
considered here, e.g., $M_H=110$ GeV (where the bottom-antibottom
channel is unrivaled by any other decay mode), this corresponds to
$E_{\mathrm{cm}}=500$ GeV. 
Furthermore, 
the sensitivity of the production rates of reaction 
(\ref{eeHHZ}) on $\lambda_{HHH}$ is higher at lower collider energies
\cite{PMZ}. Thus, in order to illustrate the interplay
between reactions (\ref{eeHHZ})--(\ref{eeQCD}), we will in the following 
focus on the case of a CM energy of 500 GeV, top-right corner
of Fig.~\ref{fig_cross}, as an 
illustrative example. In fact, the discussion for the other two choices,
$\Ecm=1000$ and 1500 GeV (the two bottom plots of Fig.~\ref{fig_cross}),
would be rather similar, so we refrain  from repeating it. 
(Also note that the signal-to-background ($S/B$)
ratio improves with increasing energy.) 

\begin{figure}[p]
~\epsfig{file=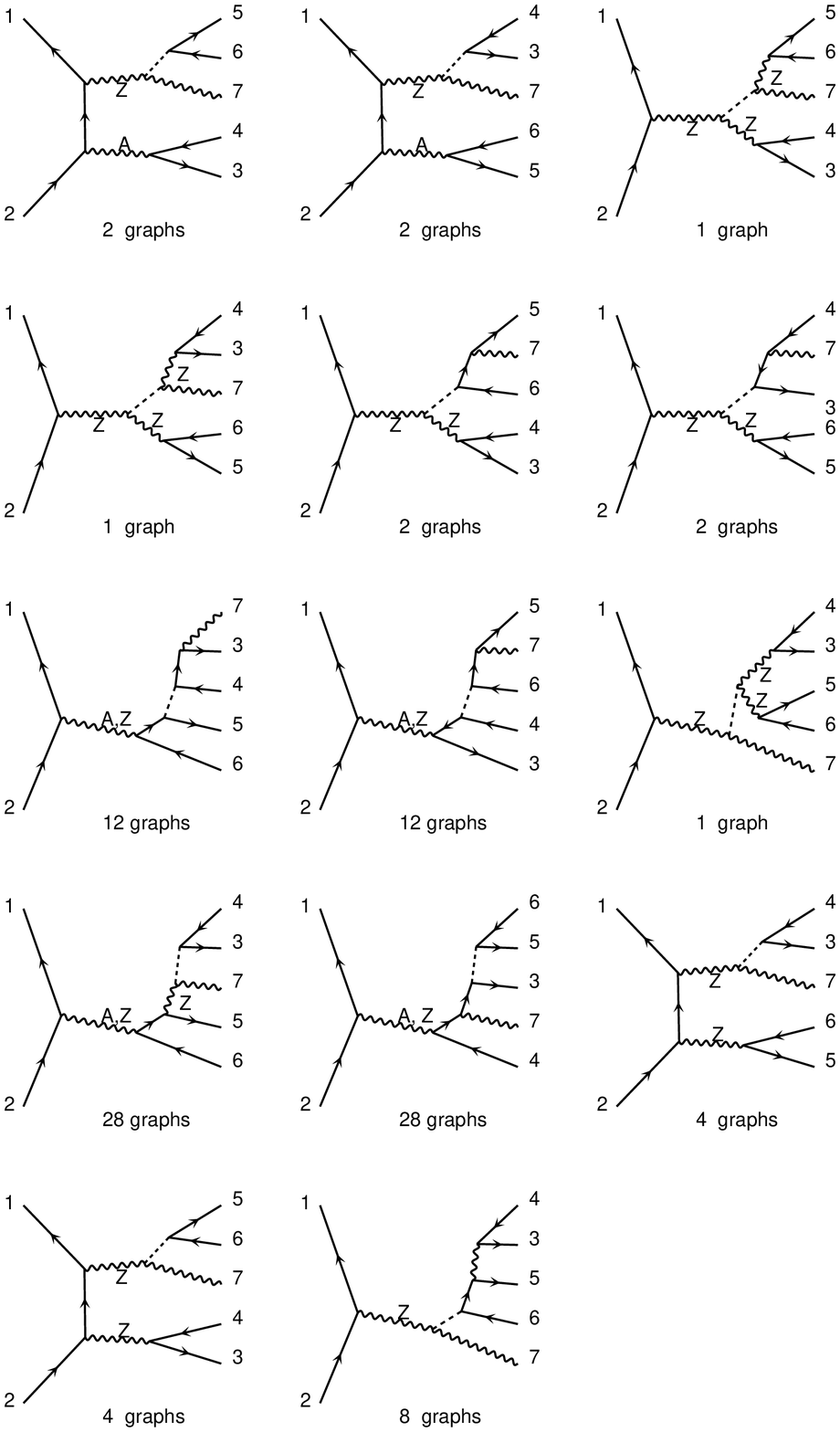,height=20cm}
\vskip-1.0cm
\caption{Topologies contributing at lowest-order to
$e^+_1e^-_2 \to b_3\bar b_4 b_5\bar b_6 Z_7$ 
via purely EW interactions
containing only one Higgs boson in intermediate states. An internal wavy line
corresponds to a $\gamma$ or a $Z$ (labeled as {\tt A} and {\tt Z},
respectively), as appropriate. The total number of actual diagrams is 107.
Finally,
diagrams which differ from 
those above only in the exchange $3\leftrightarrow5$ 
(or, equivalently, $4\leftrightarrow6$) must also be considered,
preceded by a minus sign.}
\label{fig_H}
\end{figure}

\begin{figure}[p]
~\epsfig{file=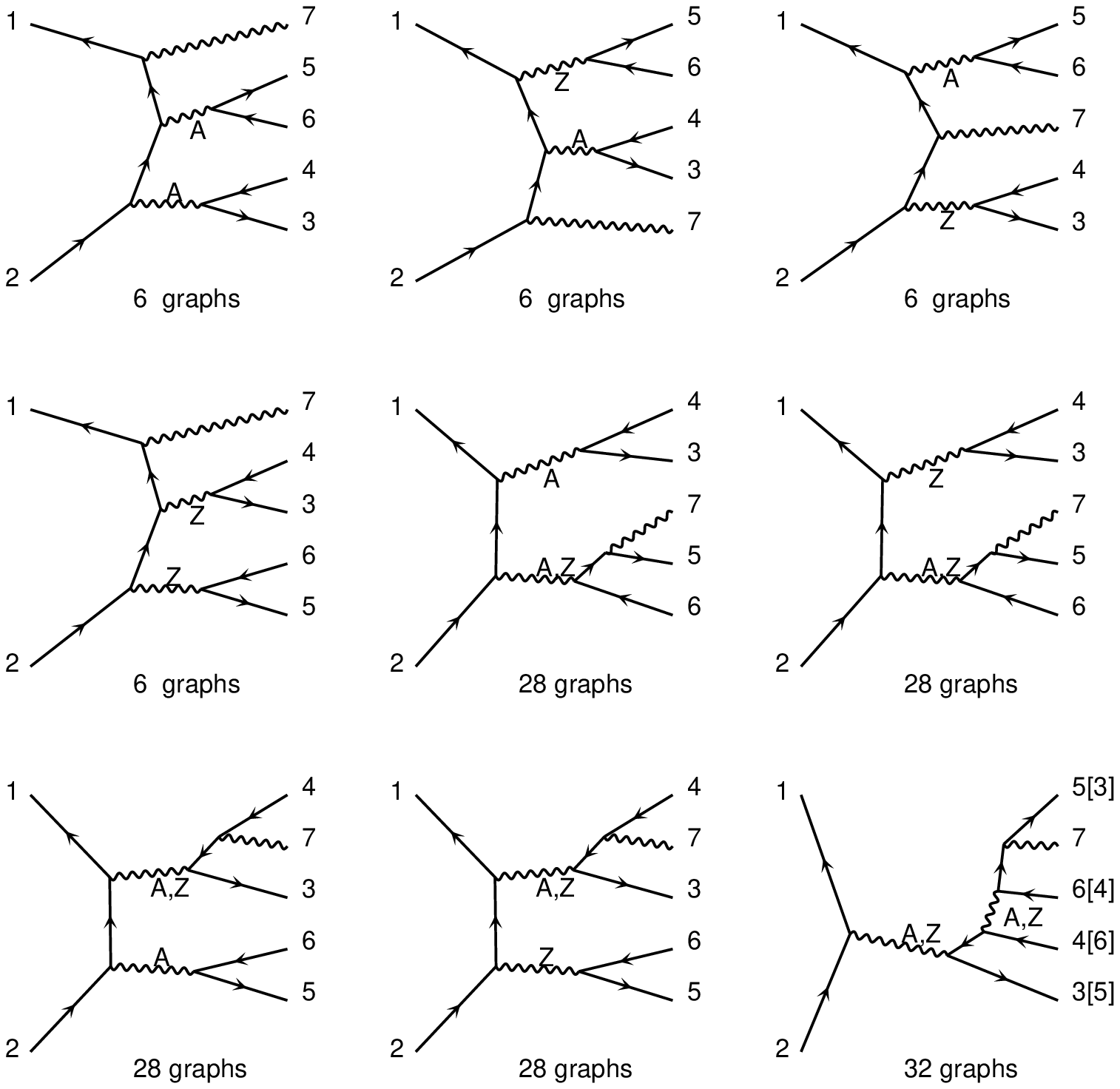,height=20cm}
\vskip-8.5cm
\caption{Topologies contributing at lowest-order to
$e^+_1e^-_2 \to b_3\bar b_4 b_5\bar b_6 Z_7$ 
via purely EW interactions
containing no Higgs bosons in intermediate states. An internal wavy line
corresponds to a $\gamma$ or a $Z$ (labeled as {\tt A} and {\tt Z},
respectively), as appropriate. The total number of actual diagrams is 168.
Finally,
diagrams which differ from 
those above only in the exchange $3\leftrightarrow5$ 
(or, equivalently, $4\leftrightarrow6$) must also be considered,
preceded by a minus sign.}
\label{fig_N}
\end{figure}

The rise at 500 GeV of the purely EW background (\ref{eeEW}) with the Higgs
mass can be understood in the following terms.
The dominant components of the EW process are those given by: 
\begin{enumerate}
\item $e^+e^-\to ZZZ\to b\bar b b\bar b Z$, 
first from the left in the second row of topologies in Fig.~\ref{fig_N}.
That is, triple $Z$ production with no Higgs boson involved.
\item $e^+e^-\to HZZ\to b\bar b b\bar b Z$,
first(first) from the left(right) 
in the fifth(fourth) row  of  topologies in Fig.~\ref{fig_H}
(also including the diagrams in which the on-shell $Z$ is connected to
the electron-positron line). 
That is,  single Higgs-strahlung 
production  in association with an additional $Z$, with the Higgs
decaying to $b\bar b$. The cross sections of these two channels are 
obviously identical.
\item $e^+e^-\to HZ\to Z^*Z^*Z\to b\bar b b\bar b Z$,
first from the right in the third row of topologies in Fig.~\ref{fig_H}.
That is, single Higgs-strahlung production with the
Higgs decaying to $b\bar b b\bar b$ via two off-shell $Z^*$ bosons. 
\item $e^+e^-\to ZH\to b\bar b Z^*Z\to b\bar b b\bar b Z$,
first(first) from the right(left) 
in the first(second) row of topologies in Fig.~\ref{fig_H}.
That is, two single Higgs-strahlung production channels with the
Higgs decaying to $b\bar b Z$ via one off-shell $Z^*$ boson. Also the
cross sections of these two channels are identical to each other, as in 2.
\end{enumerate}
The production rates of 1.--4. as separate subprocesses can be found in 
the upper portion of Fig.~\ref{fig_split}.
All other EW subprocesses
are much smaller and rarely exceed $10^{-3}$ femtobarns,
so we do not plot them here.

\begin{figure}[p]
~\epsfig{file=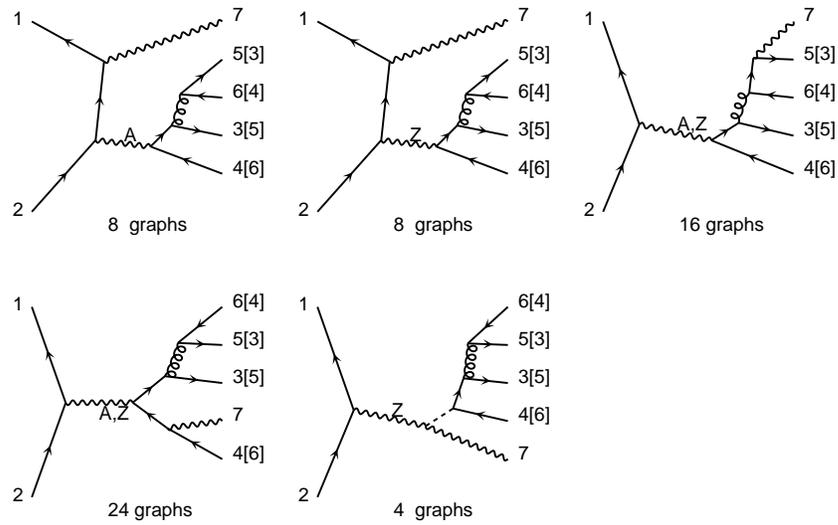,height=20cm}
\vskip-12.5cm
\caption{Topologies contributing at lowest-order to
$e^+_1e^-_2 \to b_3\bar b_4 b_5\bar b_6 Z_7$ 
via  QCD interactions
containing one gluon in intermediate states. An internal wavy line
corresponds to a $\gamma$ or a $Z$ (labeled as {\tt A} and {\tt Z},
respectively), as appropriate, whereas a helical one refers to a $g$.
The total number of actual diagrams is 60. Finally,
diagrams which differ from 
those above only in the exchange $3\leftrightarrow5$ 
(or, equivalently, $4\leftrightarrow6$) must also be considered,
preceded by a minus sign.}
\label{fig_Q}
\end{figure}

The QCD process (\ref{eeQCD}) is dominated by $e^+e^-\ar ZZ$ 
production with one
of the two $Z$ bosons decaying hadronically into four $b$ jets.
This subprocess corresponds to the topology in the middle of the
first row of diagrams in Fig.~\ref{fig_Q}. Notice that Higgs
diagrams are involved in this process as well (bottom-right topology in 
the above figure). These correspond to single Higgs-strahlung production
with the Higgs scalar subsequently decaying into $b\bar bb\bar b$
via an off-shell gluon. Their contribution is not negligible, owning to
the large $ZH$ production rates, as can be seen in 
the lower portion of Fig.~\ref{fig_split}. The somewhat unexpected
dependence of the latter upon $M_H$ (with a maximum at 130 GeV) is
the result of the interplay between our acceptance cuts and phase space 
effects. The contribution of the other diagrams, which do not resonate,
is typically one order of magnitude smaller than the $ZZ$ and $ZH$ 
mediated graphs, with the interferences even smaller (and generally
negative).  

One should note from 
Fig.~\ref{fig_cross} that the overall rates of the signal
are quite small (also recall that we neglect tagging efficiency as well
as the $Z$ decay rates), even at low Higgs masses where both the production
and decay rates are largest. In fact, they are always below 
$0.2$ femtobarns for all energies from 500 to 1500 GeV, although this
can be doubled simply by polarizing the incoming electron and positron beams.
Thus, as already recognised in Ref.~\cite{PMZ}, where on-shell
production studies of process (\ref{eeHHZ}) were performed, luminosities
of the order of one inverse attobarn need to be collected
before statistically significant measurements of $\lambda_{HHH}$
can be performed. This emphasizes the need of high luminosity at any future LC.

We now proceed by looking at several differential spectra
of reactions (\ref{eeHHZ})--(\ref{eeQCD}), in order
to find suitable kinematic cuts which will enhance the 
$S/B$ ratio.
The distributions in $E(b)$ and $\cos(b,b)$ leave little to exploit in
separating signal from background after the acceptance cuts are made,
especially with respect to the EW background. 
We turn then to other spectra, for example, invariant masses of $b$ 
(anti)quark systems. In this respect, we have plotted those of the
following combinations:
\begin{itemize}
\item[{(a,b)}] of $2b$ systems, for the case in which
the $b$ jets come from the same production vertex (`right' pairing)
and the opposite case as well (`wrong' pairing);
\item[{(c)~~}]  of $3b$ systems, in which only two $b$
(anti)quarks have the same EM charge;
\item[{(d)~~}]  of the $4b$ system.
\end{itemize}
We denote the  mass spectra of the systems
(a)--(d) as $M_R(bb)$ and $M_W(bb)$
(where $R(W)$ signifies the right(wrong) combination),  $M(bbb)$ and 
$M(bbbb)$, respectively.
In the first three cases, there exists more than one combination
of $b$ quarks. In such instances, we bin them all 
in the same distribution each with identical event weight.
Further notice that the $2b$ invariant masses that can be reconstructed
experimentally are actually 
appropriately weighted superpositions of $M_R(bb)$ and $M_W(bb)$.
In particular, if the $b$ charge tag is available, then it is roughly the sum
of the two. If not, the latter is about twice as large as the former.

The invariant mass spectra can be found in 
Fig.~\ref{fig_masses}, for the combination $E_{\mathrm{cm}}=500$ GeV
and $M_H=110$ GeV. Here, one can appreciate the narrow Higgs 
peak\footnote{Recall that for $M_H=110$ GeV one has $\Gamma_H\approx3$ MeV.
The Higgs resonances in the top-left frame of Fig.~\ref{fig_masses}
have been smeared out by incorporating a $5$ GeV bin width, 
emulating the finite
efficiency of the detectors in determining energies and angles.}
in the $M_R(bb)$ distribution, that can certainly be exploited in the
signal selection, especially against the QCD background, which is rather
flat in the vicinity of $M_H$. Apparently, this is no longer true for
the EW process, as it also displays a resonance at $M_H$ (induced 
by the diagrams in Fig.~\ref{fig_H} which carry an external on-shell
current $H\to b\bar b$). However, events of the type (\ref{eeHHZ}) contain two
$2b$ invariant masses naturally peaking at $M_H$, 
whereas only one would appear in samples produced by process (\ref{eeEW}) 
(apart from accidental resonant mispairings).
Thus, even in the case of the  EW background one can achieve a
significant noise reduction.
Finally, requiring that none of the $2b$ invariant masses reproduce
a $Z$ boson will also be helpful in this respect, as evident from the $M_R(bb)$
spectrum of process (\ref{eeEW}). However, in this case, the invariant mass 
resolution of di-jet systems must be at least as good as the difference 
$(M_H-M_Z)/2$, in order to resolve the $Z$ and $H$  peaks.
Other mass distributions can be useful too in reducing
the noise while keeping a substantial portion of the signal.
Of some help are the $M(bbb)$ and $M(bbbb)$ spectra. In
particular, notice that the
minimum value of the latter is about $2M_H$ for process (\ref{eeHHZ}),
whereas for reaction (\ref{eeEW}) it is lower, typically
around $2M_Z$ or $M_H+M_Z$, as driven by the two dominant components
of the EW background at low Higgs mass (i.e., 
subprocesses 1. and 2., respectively, see top of Fig.~\ref{fig_split}).
The QCD background can stretch to $M(bbbb)$ values even further below
the $2M_H$ end point of the signal (the more the larger $M_H$), indeed
showing a peak both at $M_H$ and $M_Z$, corresponding to
the $H\to 4b$ and $Z\to 4b$ decays induced by the second and last 
topologies in Fig.~\ref{fig_Q}. As for the $M(bbb)$ spectrum, its
shape is strongly related to that of the $4b$ mass. In a sense,
by excluding one of the four $b$ quarks from the mass reconstruction
corresponds to smearing the $M(bbbb)$ distributions, so that the 
broad prominent peak at $M(bbb)\approx90-100$ GeV in the case of the QCD
process
can be viewed as the superposition of the remains of the two narrow
ones seen in $M(bbbb)$.
For this very reason then, once a selection cut is imposed on
one of the two masses, this is very likely to affect the other in a similar
manner.

The spectacular differences seen in Fig.~\ref{fig_masses} between,
on the one hand, process (\ref{eeHHZ}), and,
on the other hand, reactions (\ref{eeEW}) and (\ref{eeQCD}) (more in the
former than in the latter), are a direct
reflection of the rather different resonant structure of the various
channels, that is, the form of the time-like propagators (i.e., $s$-channels)
in the corresponding
MEs. However, one should expect further kinematic differences, driven
by the presence in the backgrounds of space-like propagators
 (i.e., $t,u$-channels), which are instead absent in the signal
(see Figs.~\ref{fig_HH}--\ref{fig_Q}). This is most evident in the properties
of the four-quark hadronic system (d) recoiling against the $Z$ boson. 
As the internal dynamics of the four $b$ quarks is very different
in each process (\ref{eeHHZ})--(\ref{eeQCD}),
 we also study the cases (a)--(c) separately.

One can appreciate the propagator effects by plotting, for example,
the cosine of the polar angle (i.e., with the beam axes) of the four $b$ quark
system (or, indeed, the real $Z$). See Fig.~\ref{fig_cos}, where, again, 
$E_{\mathrm{cm}}=500$ GeV and $M_H=110$ GeV. 
Notice that the backgrounds are much more forward peaked than the signal.
This can be understood by recalling that the QCD events
are mainly due to $e^+e^-\to ZZ$ production
followed by the decay of one of the gauge bosons into four $b$ quarks.
The $ZZ$ pair is produced via $t,u$-channel graphs, so the
gauge bosons are preferentially directed forward and backward into the
detector. In contrast,
the signal (\ref{eeHHZ}) is always induced by $s$-channel graphs.
The EW background (\ref{eeEW}) has a more complicated structure but is still
sizably dominated by forward production.  The behaviour of
$\cos(bbb)$ and $\cos(bb)$ is very similar to that of the four-quark
system.
In practise, it is the strong boost of the $Z$ bosons
produced forwards and backwards in processes (\ref{eeEW})--(\ref{eeQCD}),
combined with the small value of $m_b$ (compared to the typical
process scale, i.e., $E_{\mathrm{cm}}$), that produces a similar angular
pattern
for all multi-$b$ systems of the background, regardless of their actual 
number.  This is true for reaction (\ref{eeHHZ}) also.
Therefore, all angular distributions displayed can boast 
strong (though correlated) discriminatory powers, allowing one to separate 
signal and backgrounds events  efficiently. 

An alternative possible means of disentangling the effects of the propagators
is to resort to the differential distributions
of the above systems (a)--(d) in transverse momentum, $pT$. These are
plotted in Fig.~\ref{fig_pTs}, for the same $\Ecm$ and $M_H$ as the previous
two figures. However, this kinematic variable proves not to be useful. In
fact, the only discriminating distribution is the $pT$ for all four $b$
quarks (or equivalently the final state $Z$ boson), and this only singles out
the QCD background, a large fraction of which
populates the  range beyond 180--200 GeV. Neither the signal
nor the EW background do so and always look rather similar
in their shape (even in the spectra involving $2b$ quarks only, once these 
are appropriately combined together).
 
Notice that two of the $b$ quarks in the QCD background originate from gluon
splitting and are therefore rather soft. Consequently, one expects the four
$b$ quarks to be more planar, in the $4b$ rest frame, for the QCD background 
than for the signal, where they are all the decay products of heavy bosons.
We study this by plotting the thrust $T$ \cite{T} and sphericity $S$ \cite{S} 
distributions in Fig.~\ref{fig_shape}. Indeed, such quantities
could prove useful in reducing the QCD background but are harmless against
the EW background.

Before proceeding to apply dedicated selection cuts,
we remark that kinematic features similar to those displayed 
in Figs.~\ref{fig_masses}--\ref{fig_shape} can also be seen at the
other two values of CM energies considered here and for other
Higgs masses. In fact,
increasing the value of $M_H$ in those distributions mainly 
translates into a `movable'
resonant peak in $M_R(bb)$ as well as lower-end point in $M(bbbb)$
and into somewhat softer(harder) spectra in invariant mass(transverse 
momentum) than at the smaller $M_H$ value considered so far. Moreover,
 the $t,u$-channel
dependence of the backgrounds, as opposed to the $s$-one of the 
signal, is more marked at higher $E_{\mathrm{cm}}$ values. Finally,
for angular distributions, a larger Higgs mass does not remove the big
differences seen between the three channels
(\ref{eeHHZ})--(\ref{eeQCD}).

Therefore, in all generality, following our discussions of
Figs.~\ref{fig_masses}--\ref{fig_shape}, and recalling the need
to  economize on the loss of signal because of its rather
small production and decay rates, one can optimise the
$S/B$ ratio by imposing the cuts:
$$
|M(bb)-M_H|<5~{\mathrm{GeV}}~({\mathrm{on~exactly~two~combinations~of}}~
2{\it b}~{\mathrm{systems}}),
$$
$$
|M(bb)-M_Z|>5~{\mathrm{GeV}}~({\mathrm{for~all~combinations~of}}~
2{\it b}~{\mathrm{systems}})
$$
\begin{equation}\label{cuts}
M(bbbb)>2M_H,
\qquad\qquad |\cos(2b,3b,4b)|<0.75.
\end{equation}
In enforcing these constraints, we assume no $b$ jet charge
determination. Moreover, the reader should 
recall that the spectra of the four
hadronic systems (a)--(d) are all correlated, in each of the quantities 
studied above, and so are the invariant masses, transverse momenta
and polar angles among themselves.

The counterpart of Fig.~\ref{fig_cross} after the implementation
of the above cuts is  Fig.~\ref{fig_cross_cut}.
 The effect of the latter is a drastic
reduction of both background rates (\ref{eeEW})--(\ref{eeQCD}),
 while maintaining a large portion of the original signal (\ref{eeHHZ}).
Further notice how the imposition of the cuts (\ref{cuts}) modifies
the hierarchy of cross sections for process (\ref{eeHHZ}) with the
CM energy, as now the largest rates occur at $\Ecm=1000$ GeV
and the smallest at $\Ecm=500$ GeV (compare to Fig.~\ref{fig_cross}).

The $S/B$ ratios turn out to be enormously large
for not too heavy Higgs masses. For example, at $\Ecm=500(1000)[1500]$
GeV and for $M_H=110$ GeV, one gets $S/B=25(60)[104]$, where $S$ corresponds
to the rates of reaction (\ref{eeHHZ}) and $B$ refers
to the sum of the cross sections for processes (\ref{eeEW})--(\ref{eeQCD}).   
The reduction of both backgrounds amounts to several order of
magnitudes, particularly for the case of the strong process,
whereas the loss of signal is much more contained.
The acceptance of the latter is better at higher collider energies
and lower Higgs masses. In fact, the poorest rate occurs for $\Ecm=
500$ GeV at the upper end of the $M_H$ range, where more than 90\%
of the signal is sacrificed.
We should however remark that the suppression of the backgrounds comes
largely from the invariant mass cuts on $M_{bb}$ advocated in 
(\ref{cuts}). (In fact, they are crucial not only in selecting
the $M_H$ resonance of the signal, but also in minimizing the
rejection of the latter around $M_Z$ when mispairings occur: notice
the shoulder at 90 GeV of the $M_W(bb)$ spectrum of reaction (\ref{eeHHZ})).
The value we have adopted for the resolution is rather high, 
 considering the large uncertainties normally associated
with the experimental
determination of jet angles and energies, though not unrealistic
in view of the most recent studies \cite{resolution}. The ability of the
actual detectors in guaranteeing the performances foreseen at present
is thus crucial for the feasibility of dedicate studies of 
double Higgs-strahlung events at the LC.

A related aspect is the efficiency of tagging the $b$ quarks 
necessarily present in the final state of reaction (\ref{eeHHZ}),
particularly in the case in which the $Z$ boson decays hadronically.
On the one hand, given the high production rate of six jet events
from QCD \cite{ee6j} and multiple gauge boson resonances \cite{gauge}
in light quark and gluon jets, it is desirable to 
resort to heavy flavour identification in hadronic samples. On the other
hand, the poor statistics of the $HHZ$ signal requires a judicious approach 
in order not to deplete the latter below detection level. According
to recent studies \cite{btag}, the two instances
can be combined successfully, as 
efficiencies for tagging $b\bar b$ pairs produced in  Higgs decays
were computed to be as large as $\epsilon_{b\bar b}\approx90\%$,
with mis-identification probabilities of light(charmed) quarks as 
low as $\epsilon_{q\bar q(c\bar c)}\approx0.3(4)\%$
(and negligible for  gluons).
If such a projection for the LC detectors
proves to be true, then even the requirement
of tagging exactly four $b$ quarks in double-Higgs events of the
type (\ref{eeHHZ}) might be statistically feasible, thus suppressing
the reducible backgrounds to really marginal levels \cite{Lutz}.

One must also bear in mind that experimental considerations, such as the
performances of detectors, the fragmentation/hadronization dynamics and a
realistic treatment of the $Z$ boson decays, are also important when
determining what cuts should be made. Such considerations are beyond the scope
of this paper, and are under study elsewhere \cite{Lutz}. 

Finally, the number of signal and backgrounds events seen per inverse attobarn
of luminosity at $\Ecm=500$, $1000$, and $1500$ GeV, with $M_H=110$ GeV, can
be seen in Tab.~\ref{eventtable}. Of course, one could relax one or more  of
the constraints we have adopted to try to improve the signal rates
without letting the backgrounds become unmanageably large. For example,
by removing the cuts on $\cos(bb)$ and $\cos(bbb)$ one
can enhance the signal rates by about a factor of two.
However, the EW background would also increase by a comparable
amount and the QCD rate would do by a somewhat larger factor, of
at least three/four. Kinematic fits can also help in improving the $S/B$
ratios \cite{Lutz}.

\section{Summary}
\label{sec_conclusions}

In conclusion, the overwhelming irreducible background 
from EW and QCD processes
of the type $e^+e^-\to b\bar b b\bar b Z$ to double Higgs 
production in association with $Z$ bosons and decay in the
channel $H\to b\bar b$, i.e.,
$e^+e^-\to HHZ\to b\bar b b\bar b Z$, should easily be suppressed down 
to manageable levels by simple kinematics cuts: e.g., in invariant masses 
and polar angles.

The number of signal events is generally rather low, but will be
observable at the LC given the following `mandatory conditions' (some of
which have already been outlined in Ref.~\cite{Lutz}):
\begin{itemize}
\item very high luminosity;
\item excellent $b$ tagging performances;
\item high di-jet resolution.
\end{itemize}
The requirement advocated in Ref.~\cite{Lutz} of a good forward acceptance
for jets may also be added to the above list, as we have
explicitly verified (though not shown) that single jet directions can stretch
in the double Higgs-strahlung process up to about 20 degrees in polar
angle. Finally, beam polarization can also be invoked to increase the
signal-to-background rates \cite{PMZ}.

\section*{Acknowledgments}

We thank Peter Zerwas for suggesting the subject of this research and
for useful discussions. DJM would also like to thank DESY for hospitality
while part of this research was carried out. Financial support from the
UK-PPARC is gratefully acknowledged by both authors.

\vfill\clearpage\thispagestyle{empty}

\begin{table}[!t]
\begin{center}
\begin{tabular}{|c||c|c|c|}
\hline
& \multicolumn{3}{c|}{Number of Events per $ab^{-1}$ after selection 
cuts}  \\ \cline{2-4}
& $E_{\rm cm}=500$ GeV & $E_{\rm cm}=1000$ GeV & $E_{\rm cm}=1500$ GeV \\ \hline\hline
signal & $26$ & $40$ & $34$ \\ \hline
Electroweak& $1.0$ & $0.6$ & $0.3$ \\ \hline
QCD & $0.032$ & $0.026$ & $0.016$ \\ \hline
\end{tabular}
\caption{The number of events for signal and backgrounds per inverse attobarn
of luminosity after selection cuts for centre-of-mass energies of $500$,
$1000$ and $1500$ GeV, a Higgs mass of $110$ GeV, and with polarized 
electron and positron beams.}
\label{eventtable}
\end{center}
\end{table}

\vfill\clearpage\thispagestyle{empty}
 
\vfill\clearpage\thispagestyle{empty}
 
\vfill\clearpage\thispagestyle{empty}

\vfill\clearpage\thispagestyle{empty}

\begin{figure}[t!]
\begin{minipage}[b]{.495\linewidth}
\centering\epsfig{file=signal.ps,angle=90,height=6cm,width=\linewidth}
\end{minipage}\hfill\hfill
\begin{minipage}[b]{.495\linewidth}
\centering\epsfig{file=background500.ps,angle=90,height=6cm,width=\linewidth}
\end{minipage}\hfill\hfill
\begin{minipage}[b]{.495\linewidth}
\centering\epsfig{file=background1000.ps,angle=90,height=6cm,width=\linewidth}
\end{minipage}\hfill\hfill
\begin{minipage}[b]{.495\linewidth}
\centering\epsfig{file=background1500.ps,angle=90,height=6cm,width=\linewidth}
\end{minipage}
\caption{Top-left: cross sections in femtobarns for the signal at three 
different collider energies: 500, 1000 and 1500 GeV. 
Top-right(Bottom-left)[Bottom-right]: cross sections in femtobarns
for the signal versus the EW and QCD backgrounds at 500(1000)[1500] GeV.
Our acceptance cuts in energy and separation of the four $b$ quarks
have been implemented.
}
\label{fig_cross}
\end{figure}

\vfill\clearpage\thispagestyle{empty}

\begin{figure}[!t]
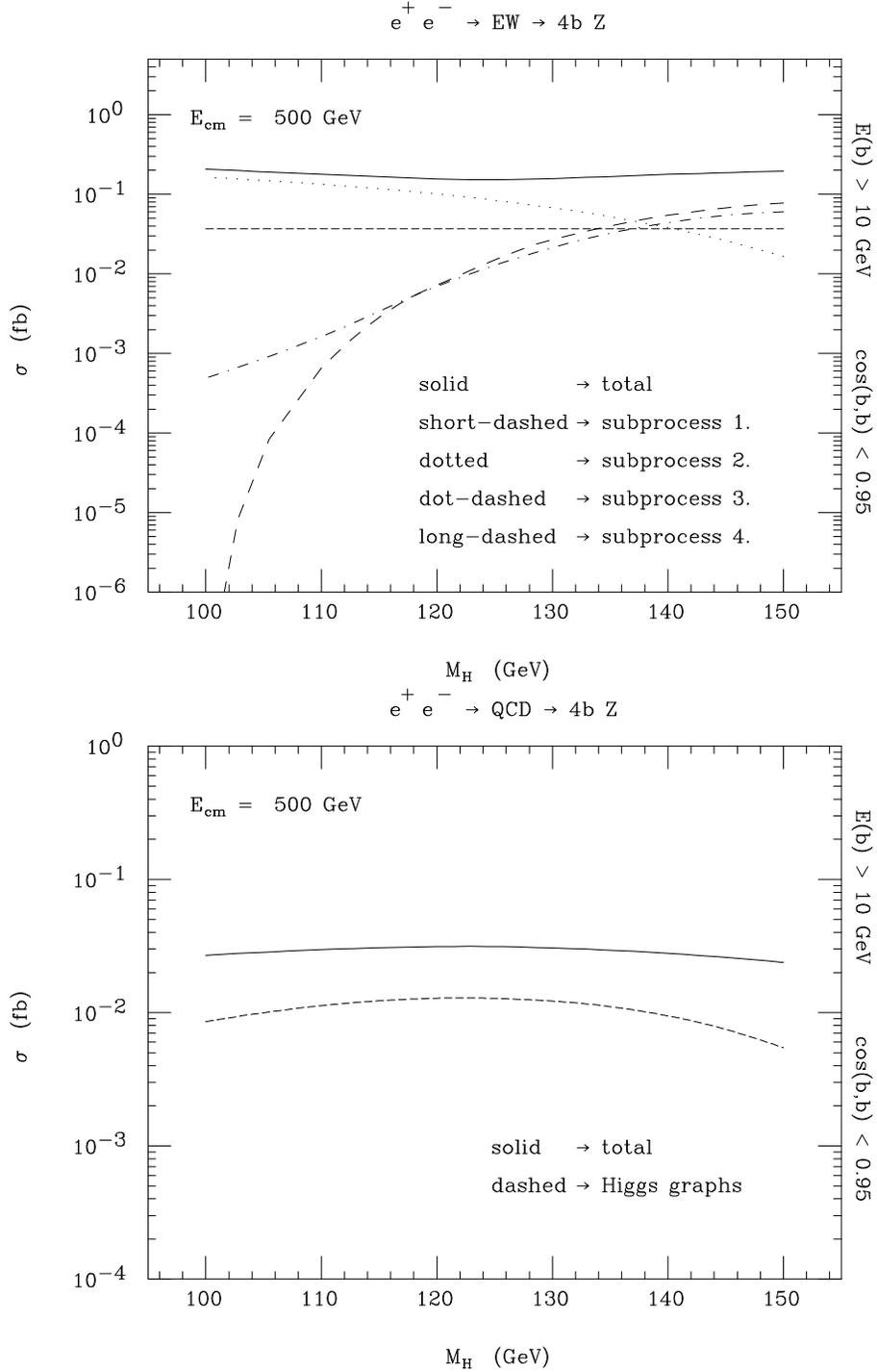

~\hskip1.5cm
{\epsfig{file=split_EW.ps,height=12cm,angle=90}}
\vskip+0.0005cm
~\hskip1.5cm
{\epsfig{file=split_QCD.ps,height=12cm,angle=90}}
\caption{Top: cross sections in femtobarns for the four
dominant components (see the text) of the purely EW background.
Bottom: cross sections in femtobarns for the total and Higgs components (see
the text) of the QCD background.
The CM energy is 500 GeV.
Our acceptance cuts in energy and separation of the four $b$ quarks
have been implemented.}
\label{fig_split}
\end{figure}

\vfill\clearpage\thispagestyle{empty}

\begin{figure}[htb]
\begin{center}
~\epsfig{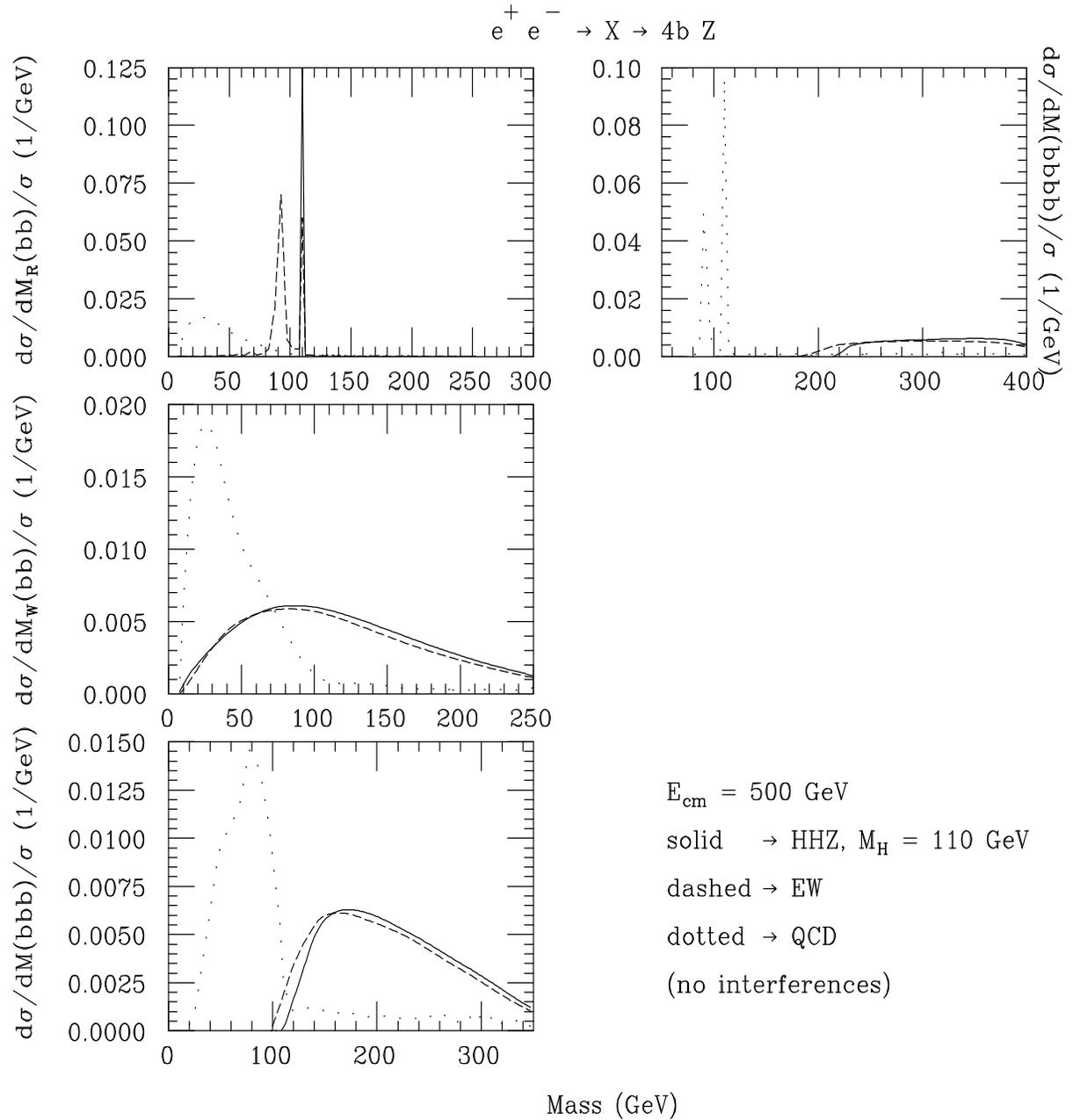}
\caption{Differential distributions in invariant mass 
of multi-jet systems containing one, two, three and four $b$ (anti)quarks.
The CM energy is 500 GeV and the Higgs mass 110 GeV.
Our acceptance cuts in energy and separation of the four $b$ quarks
have been implemented. 
}
\label{fig_masses}
\end{center}
\end{figure}

\vfill\clearpage\thispagestyle{empty}

\begin{figure}[htb]
\begin{center}
~\epsfig{file=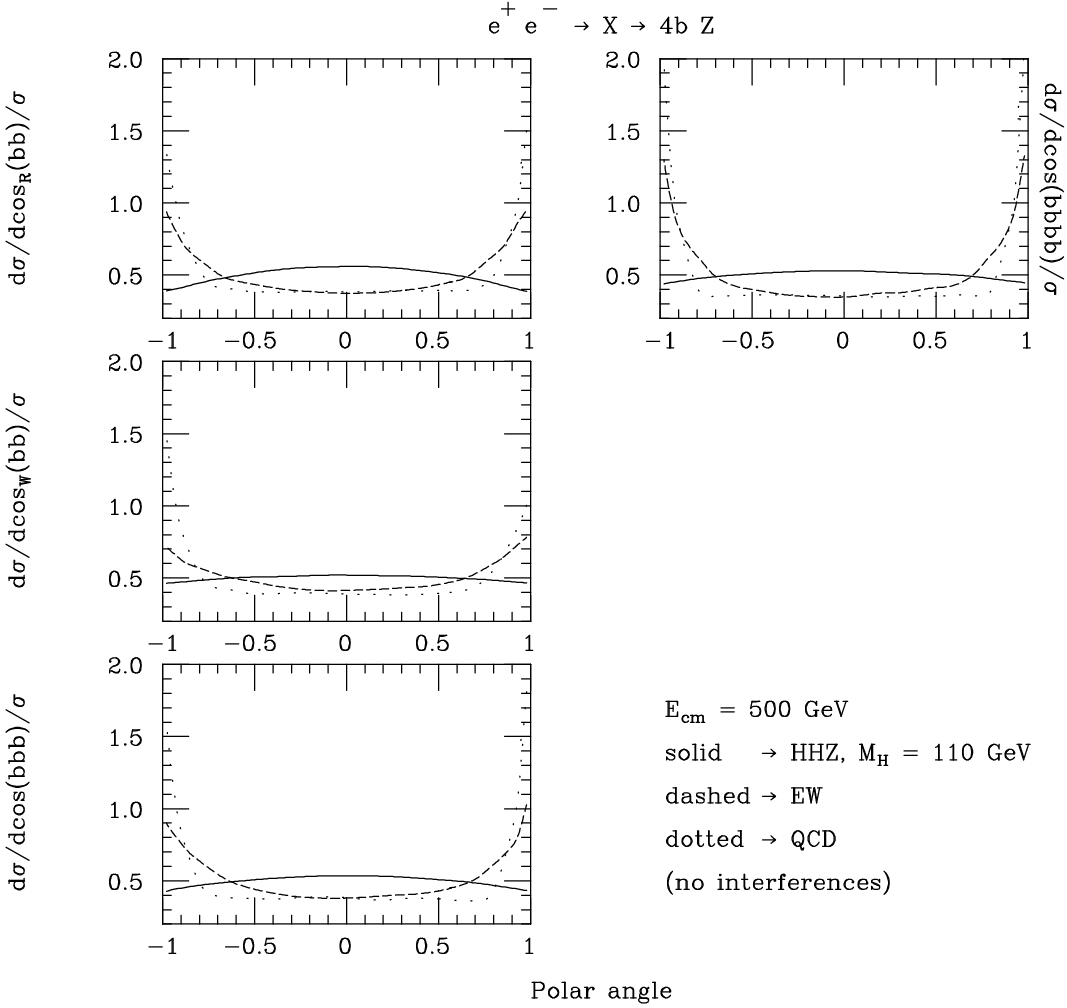,width=16cm,height=17cm,angle=0}
\caption{Differential distributions in (cosine of the) polar angle
of multi-jet systems containing one, two, three and four $b$ (anti)quarks.
(The spectrum of the $4b$ system is identical to that of the $Z$ boson.) 
The CM energy is 500 GeV and the Higgs mass 110 GeV.
Our acceptance cuts in energy and separation of the four $b$ quarks
have been implemented. 
}
\label{fig_cos}
\end{center}
\end{figure}

\vfill\clearpage\thispagestyle{empty}

\begin{figure}[htb]
\begin{center}
~\epsfig{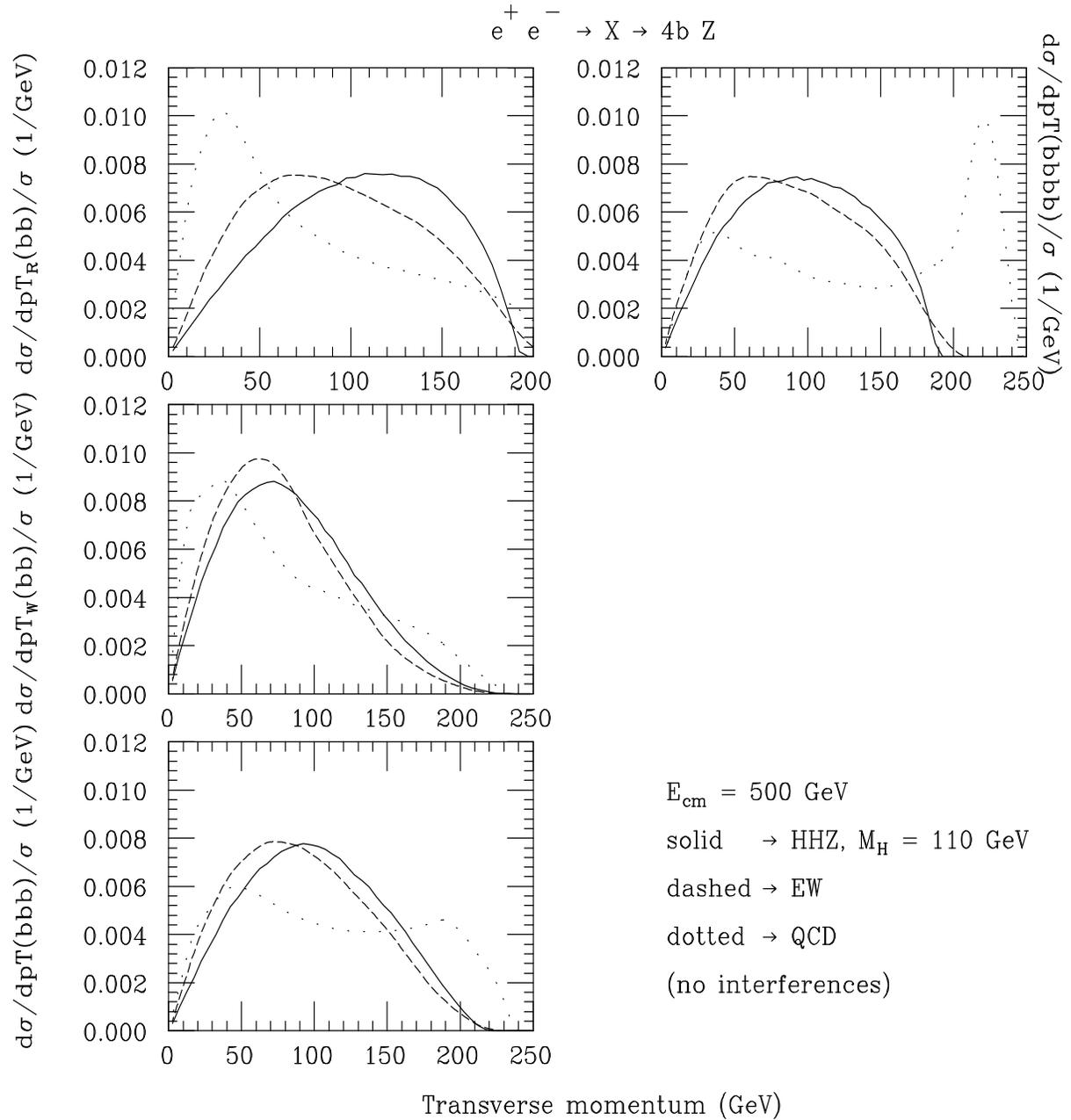}
\caption{Differential distributions in transverse momentum
of multi-jet systems containing one, two, three and four $b$ (anti)quarks.
(The spectrum of the $4b$ system is identical to that of the $Z$ boson.) 
The CM energy is 500 GeV and the Higgs mass 110 GeV.
Our acceptance cuts in energy and separation of the four $b$ quarks
have been implemented. 
}
\label{fig_pTs}
\end{center}
\end{figure}

\vfill\clearpage\thispagestyle{empty}

\begin{figure}[htb]
\begin{center}
~\epsfig{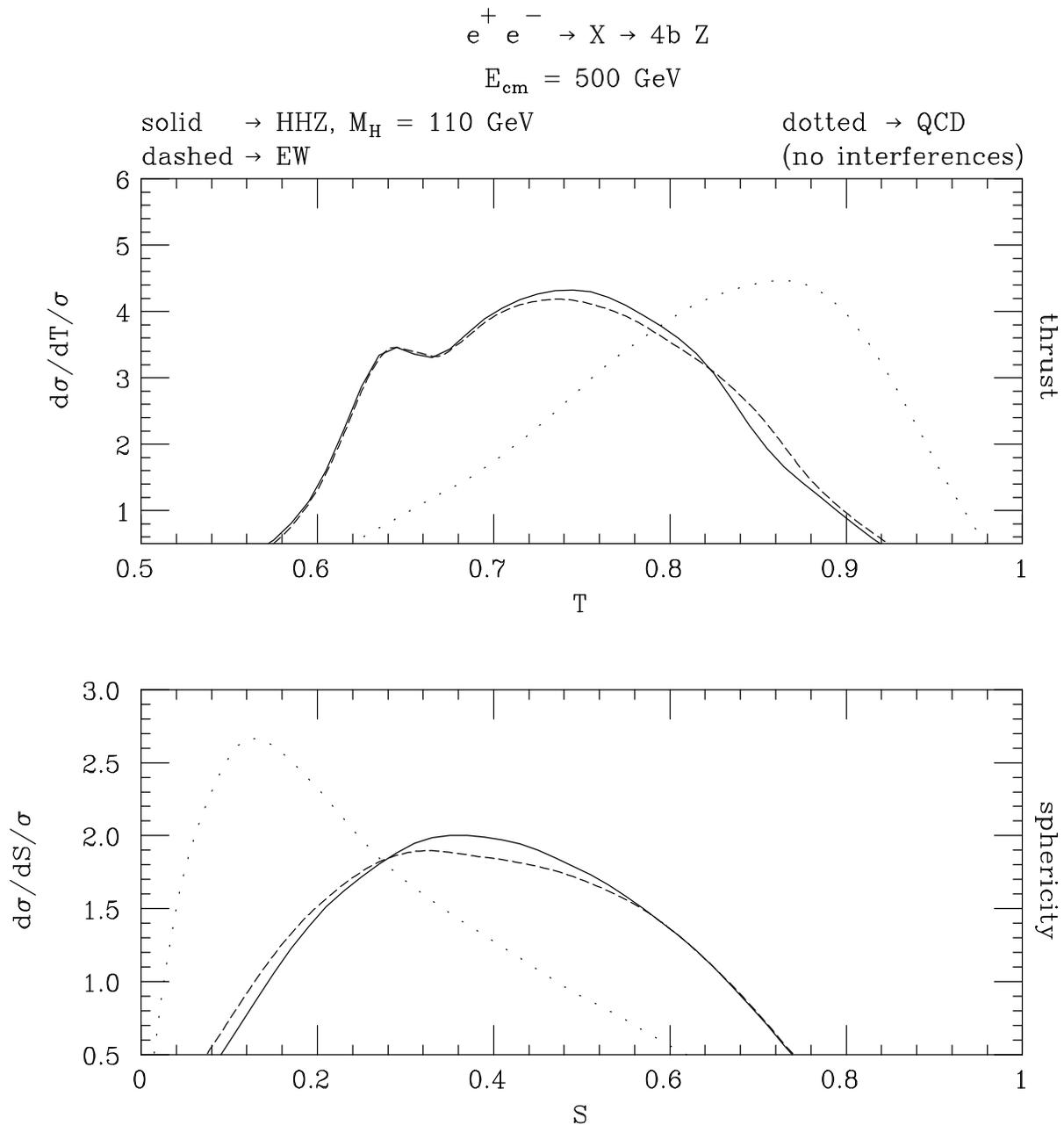}
\caption{Differential distributions in thrust and
sphericity.
The CM energy is 500 GeV and the Higgs mass 110 GeV.
Our acceptance cuts in energy and separation of the four $b$ quarks
have been implemented. 
}
\label{fig_shape}
\end{center}
\end{figure}

\vfill\clearpage\thispagestyle{empty}

\begin{figure}[t!]
\begin{minipage}[b]{.495\linewidth}
\centering\epsfig{file=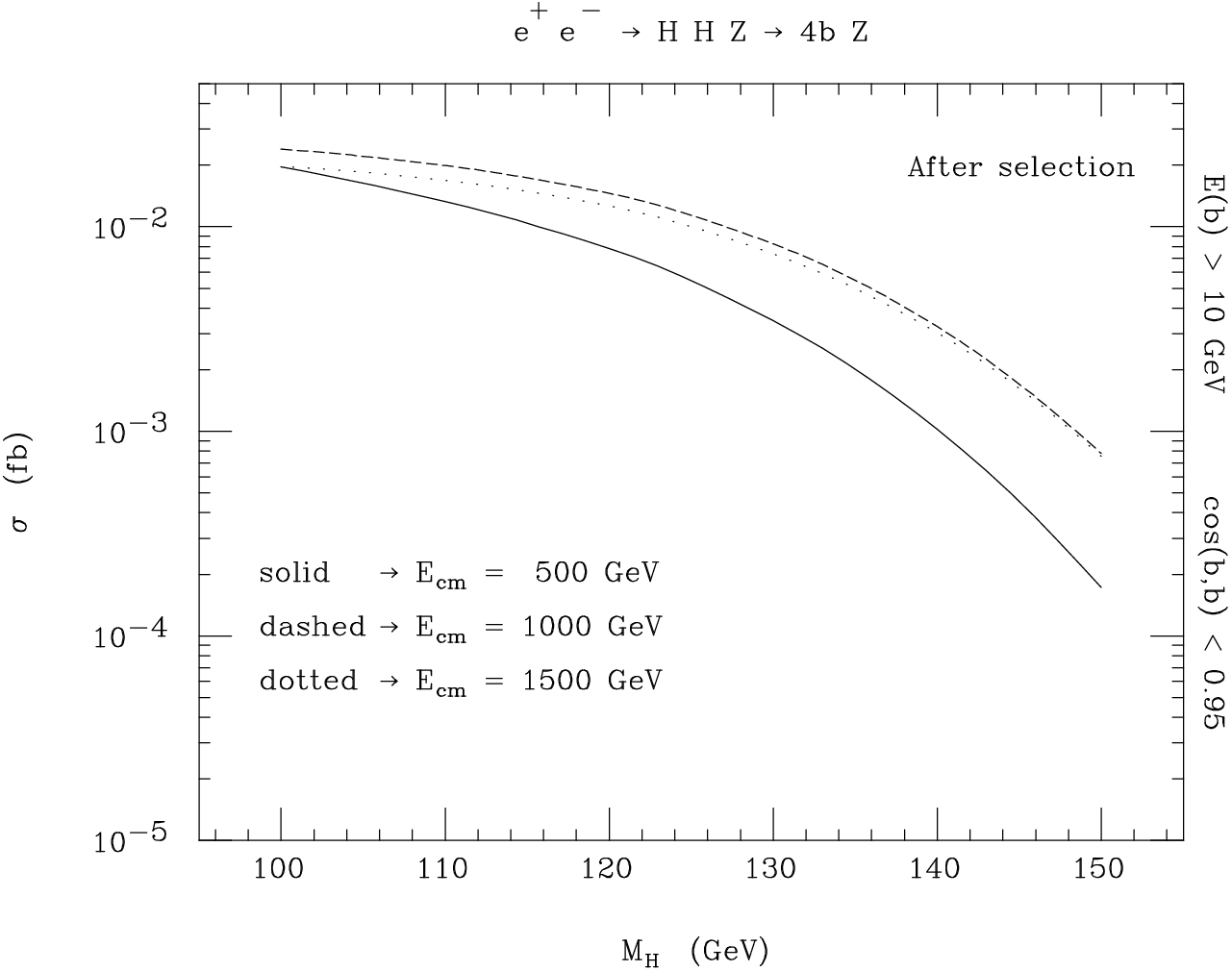,angle=90,height=6cm,width=\linewidth}
\end{minipage}\hfill\hfill
\begin{minipage}[b]{.495\linewidth}
\centering\epsfig{file=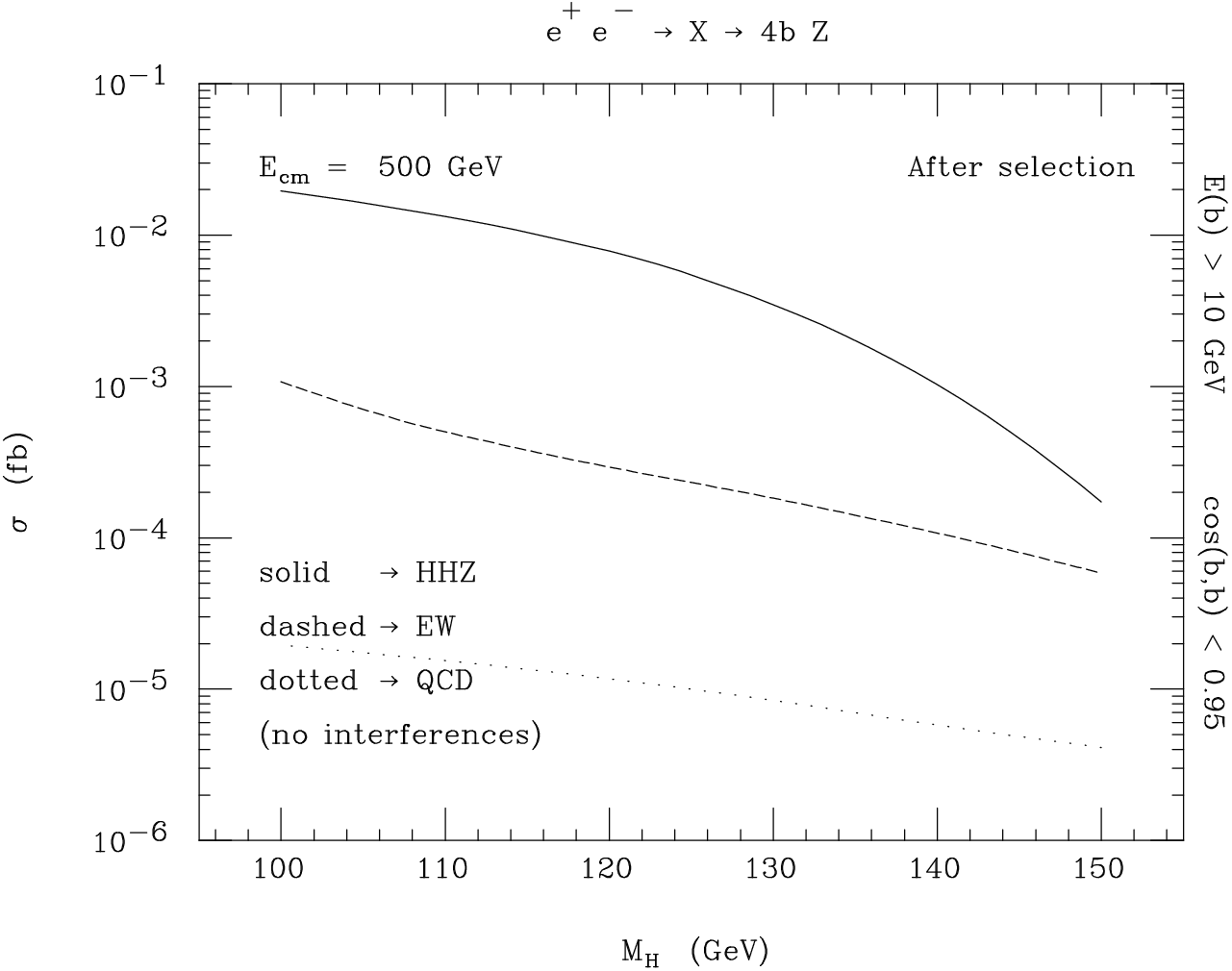,angle=90,height=6cm,width=\linewidth}
\end{minipage}\hfill\hfill
\begin{minipage}[b]{.495\linewidth}
\centering\epsfig{file=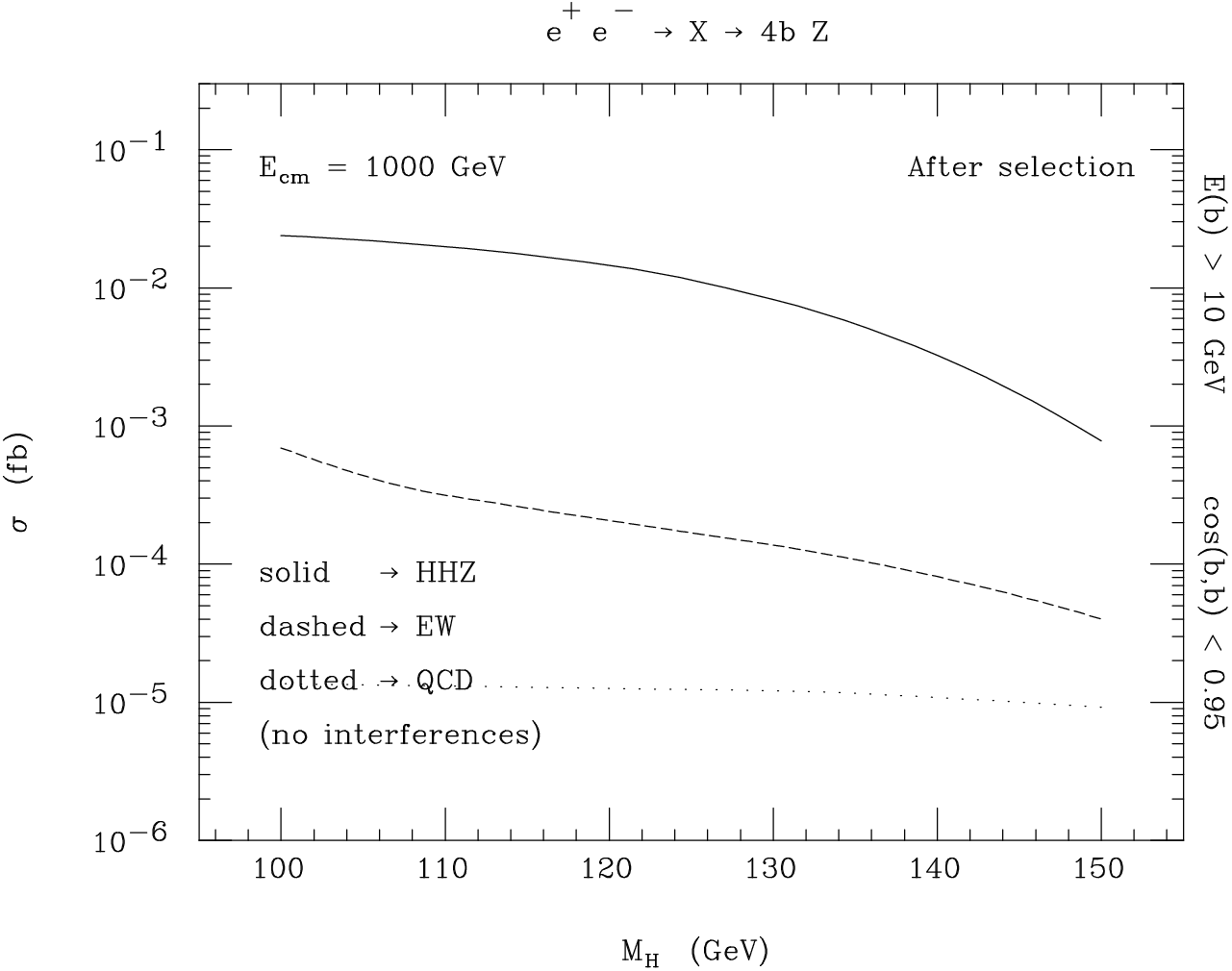,angle=90,height=6cm,width=\linewidth}
\end{minipage}\hfill\hfill
\begin{minipage}[b]{.495\linewidth}
\centering\epsfig{file=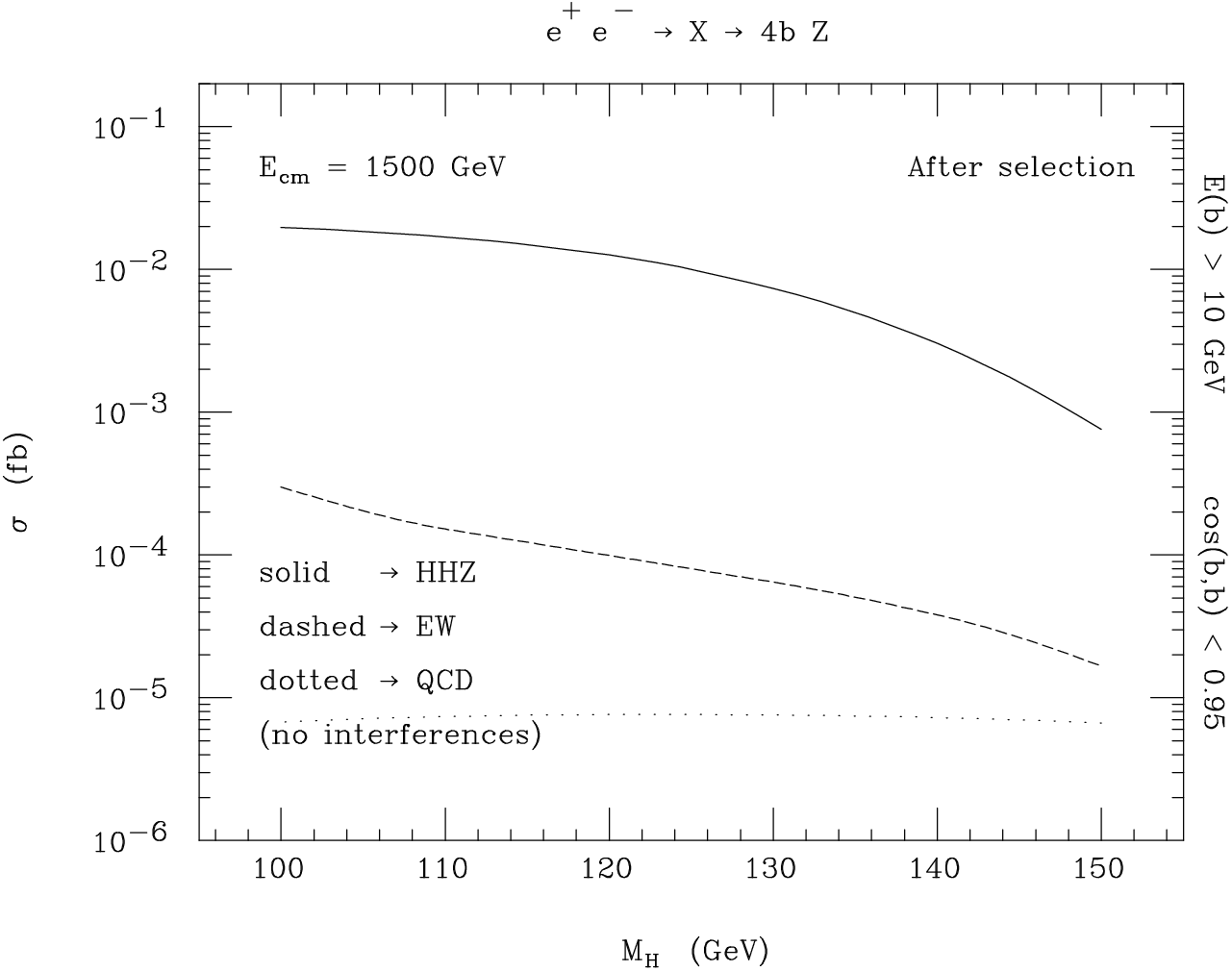,angle=90,height=6cm,width=\linewidth}
\end{minipage}
\caption{Top-left: cross sections in femtobarns for the signal at three 
different collider energies: 500, 1000 and 1500 GeV. 
Top-right(Bottom-left)[Bottom-right]: cross sections in femtobarns
for the signal versus the EW and QCD backgrounds at 500(1000)[1500] GeV.
Our acceptance cuts in energy and separation of the four $b$ quarks
have been implemented along with the selection cuts of eq.~(\ref{cuts}).
}
\label{fig_cross_cut}
\end{figure}


\begin{thebibliography}{99}

\bibitem{ee500} Proceedings of the Workshop {\it $e^+e^-$ Collisions at
500 GeV. The Physics Potential},
Munich, Annecy, Hamburg, 3--4 February 1991, ed.~P.M.~Zerwas, DESY 92--123A/B,
August 1992, DESY 93--123C, December 1993. \\
E. Accomando et al., {\it Phys. Rep. Vol. 299}, {\bf No. 1} (1998) 1.

\bibitem{revZHH} G.~Gounaris, D.~Schildknecht and F.~Renard, {\it Phys.~Lett.}
{\bf B83} (1979) 191; Erratum, {\it ibidem} {\bf B89} (1980) 437; \\
V.~Barger, T.~Han and R.J.N.~Phillips, {\it Phys.~Rev.} {\bf D38} (1988) 2766.

\bibitem{IPKSK} V.A.~Ilyin, A.E.~Pukhov, Y.~Kurihara, Y.~Shimizu and
T.~Kaneko, {\it Phys.~Rev.} {\bf D54} (1996) 6717.

\bibitem{BC} F.~Boudjema and E.~Chopin, {\it Z.~Phys.} {\bf C71} (1996) 431.

\bibitem{revWW} V.~Barger and T.~Han, {\it Mod.~Phys.~Lett.} {\bf A5} (1990)
667; \\
A.~Dobrovolskaya and V.~Novikov, {\it Z.~Phys.} {\bf C52} (1991) 427; \\
D.A.~Dicus, K.J.~Kallianpur and S.S.D.~Willenbrock, {\it Phys.~Lett.} {\bf
B200} (1998) 187; \\
A.~Abbasabadi, W.W.~Repko, D.A.~Dicus and R.~Vega, {\it Phys.~Rev.} {\bf D38}
(1998) 2770; {\it Phys.~Lett.} {\bf B213} (1998) 386.

\bibitem{revgg} E.W.N.~Glover and J.J.~van~der~Bij, 
{\it Nucl.~Phys.}~{\bf B309}
(1988) 282; \\
T.~Plehn, M.~Spira and P.M.~Zerwas, {\it Nucl. Phys.} {\bf B479} (1996) 46; 
Erratum, {\it ibidem} {\bf B531} (1998) 655; \\
S.~Dawson, S.~Dittmaier and M.~Spira, {\it Phys. Rev.} {\bf D58} (1998) 115012.

\bibitem{jikia} G.~Jikia, {\it Nucl.~Phys.} {\bf B412} (1994) 57.

\bibitem{revMSSM}
A.~Djouadi, H.E.~Haber and P.M.~Zerwas, {\it Phys.~Lett.} {\bf B375} (1996)
203; \\
P.~Osland and P.N.~Pandita, {\it Phys.~Rev.} {\bf D59} (1999) 055013; 
preprint BERGEN-1999-01, February 1999, {\tt
hep-ph}{\tt /9902270}; \\
P.~Osland, preprint ISSN 0803-2696, March 1999, {\tt hep-ph/9903301}.

\bibitem{PMZ} A.~Djouadi, W.~Kilian, M.~Muhlleitner and P.M.~Zerwas,
preprint DESY 99/001, TTP99-02, PM/99-01, March 1999,
{\tt hep-ph/9903229}; contribution
to the XXIX {\em International 
Conference on High Energy Physics}, Vancouver 1998, Heidelberg Report
HD-THEP 98-29.

\bibitem{jan} D.J. Miller and S. Moretti, in preparation.

\bibitem{Lutz} P. Lutz, talk delivered at the
ECFA/DESY Workshop on ``Physics and Detectors for
                      a Linear Collider'', Oxford, UK, March 20--23, 1999.

\bibitem{paps} A.~Ballestrero, E.~Maina and S.~Moretti, 
{\it Phys.~Lett.} {\bf B335} (1994) 460; \\
S.~Moretti, {\it Phys.~Rev.} {\bf D50} (1995) 6316; 
{\it Z.~Phys.} {\bf C73} (1997) 653; \\
S.~Moretti and K.~Odagiri, {\it Eur.~Phys.~J.} {\bf C1} (1998) 633.

\bibitem{HELAS} H.~Murayama, I.~Watanabe and K.~Hagiwara, HELAS: HELicity
                Amplitude Subroutines for Feynman Diagram Evaluations,
                {\it KEK Report} 91-11, January 1992.

\bibitem{VEGAS} G.P.~Lepage, {\it Jour.~Comp.~Phys.} {\bf 27} (1978) 192.

\bibitem{WJSZK}
Z.~Kunszt, S.~ Moretti and W.J.~ Stirling, 
{\it Z.~Phys.} {\bf C74} (1997) 479.

\bibitem{ISR} T.~Barklow, P.~Chen and W.~Kozanecki, in Ref.~\cite{ee500}, 
part A.

\bibitem{T} E. Fahri, \prl 39 1977 1587.

\bibitem{S} J.D. Bjorken and S.J. Brodsky, \pr D1 1970 1416.

\bibitem{resolution} F. Richard, private communication.

\bibitem{ee6j} S. Moretti, {\it Phys. Lett.} {\bf B420} (1998) 367;
{\it Nucl. Phys.} {\bf B544} (1999) 289.

\bibitem{gauge} A. Aeppli {\it et. al.}, in Ref.~\cite{ee500}, part A;\\
V. Barger and T. Han, \pl B212 1988 117;\\
V. Barger, T. Han and R.J.N. Phillips, \pr D39 1889 146;\\
A. Tofighi-Niaki and J.F. Gunion, {\it Phys. Rev.} {\bf D39} (1989) 720.

\bibitem{btag} G. Borissov, talk delivered at the
ECFA/DESY Workshop on ``Physics and Detectors for
                      a Linear Collider'', Oxford, UK, March 20--23, 1999;\\
M. Battaglia, {\it ibidem}.



\end{thebibliography}
\end{document}